\documentclass{aa}  
\usepackage{graphicx,natbib}
\usepackage{multirow}
\usepackage{epsfig}
\usepackage{rotating}
\usepackage{amsmath,amssymb,color,latexsym,longtable,array}

\usepackage{txfonts}
\bibpunct{(}{)}{;}{a}{}{,}

\begin{document}

   \title{\object{3C~294} revisited: Deep Large Binocular Telescope AO NIR images and optical spectroscopy\thanks{The LBT is an international collaboration among institutions in the United States, Italy, and Germany. LBT Corporation partners are: The University of Arizona on behalf of the Arizona Board of Regents; Istituto Nazionale di Astrofisica, Italy; LBT Beteiligungsgesellschaft, Germany, representing the Max-Planck Society, The Leibniz Institute for Astrophysics Potsdam, and Heidelberg University; The Ohio State University, and The Research Corporation, on behalf of The University of Notre Dame, University of Minnesota, and University of Virginia.}}

   \author{J. Heidt\inst{1} \and
          A. Quirrenbach\inst{1} \and
          N. Hoyer\inst{1} \and
          D. Thompson\inst{2} \and
          A. Pramskiy\inst{1} \and
          G. Agapito\inst{3} \and
          S. Esposito\inst{3} \and
          R. Gredel\inst{4} \and
          D. Miller\inst{2} \and
          E. Pinna\inst{3} \and
          A. Puglisi\inst{3} \and
          F. Rossi\inst{3} \and
          W. Seifert\inst{1} \and
          G. Taylor\inst{3}    
          }

   \institute{Landessternwarte, Zentrum f\"ur Astronomie der Universit\"at Heidelberg, K\"onigstuhl 12, 69117 Heidelberg, Germany\\
   \email{jheidt@lsw.uni-heidelberg.de}
           \and
           LBT Observatory, University of Arizona, 933 N.~Cherry Ave, Tucson, USA
             \and
             Arcetri Astrophysical Observatory, Largo E. Fermi 5, 50125 Firence, Italy
             \and
             Max-Planck-Institut f{\"u}r Astronomie, K{\"o}nigstuhl 17, 69117 Heidelberg, Germany
             }

   \date{Received NN; accepted NN}

 
  \abstract
   {High redshift radio galaxies are among the most massive galaxies at their redshift,
     are often found at the center of protoclusters of galaxies, and are expected to evolve
     into the present day massive central cluster galaxies. Thus they are a useful tool to
     explore structure formation in the young Universe.
   }
   {\object{3C~294} is a powerful FR II type radio galaxy at z = 1.786. Past studies have
     identified a clumpy structure, possibly indicative of a merging system, as well as
     tentative evidence that \object{3C~294} hosts a dual active galactic nucleus (AGN).
     Due to its proximity to a bright star,
   it has been subject to various adaptive optics imaging studies. 
   }
   {In order to distinguish between the various scenarios for \object{3C~294,} we performed deep,
     high-resolution 
     adaptive optics near-infrared imaging and optical spectroscopy of \object{3C~294} with the
     Large Binocular Telescope.
   }
   {We resolve the \object{3C~294} system into three distinct components separated by a few tenths of an
   arcsecond on our images. One is compact, the other two are extended, and all appear to be non-stellar.
   The nature of each component is unclear. The two extended components could be a galaxy with an internal
   absorption feature, a galaxy merger, or two galaxies at different redshifts. We can now uniquely associate
   the radio source of \object{3C~294} with one of the extended components. Based on our spectroscopy, we
   determined a redshift of z = 1.784$\pm$0.001, which is similar to the one previously cited.
   In addition we found a previously unreported emission line at $\lambda$6749.4 \AA\ in our spectra. It
   is not clear that it originates from \object{3C~294}. It could be the Ne $[$IV$]$ doublet $\lambda$2424/2426 \AA\
   at z = 1.783, or belong to the compact component at a redshift of z $\sim$ 4.56. We thus cannot unambiguously
   determine whether \object{3C~294} hosts a dual AGN or a projected pair of AGNs.
   }
{}

   \keywords{Instrumentation: adaptive optics -- Galaxies: active -- Galaxies: high-redshift -- (Galaxies:) quasars: emission lines -- (Galaxies:) quasars: individual: \object{3C~294}}

   \maketitle
   %

\section{Introduction}

Within the standard unified scheme for active galactic nuclei (AGN), radio galaxies
are seen at relatively large inclination angles between the observer and the jet 
axis \citep{1995PASP..107..803U}. Since most of the optical emission from the central engine is 
shielded by the dusty torus, they can be used to study their host galaxies and 
immediate environment in better detail than their low-inclination counterparts.
This is particularly important for powerful radio galaxies at high redshifts, 
which are among the most massive galaxies at their redshift \citep{2009ApJ...704..548O}.
They are expected to evolve to present day massive central 
cluster galaxies and are  often found at the center of (proto)clusters
(see \citet{2008A&ARv..15...67M} for a review). Recently, even a radio galaxy at 
z = 5.72 close to the presumed end of the epoch of reionization has been found by
\citet{2018MNRAS.480.2733S}. Thus studies of these species allow us to 
investigate the early formation of massive structures in the young Universe.

\object{3C~294} is a powerful Fanaroff-Riley type II (FRII) radio source at z = 1.786.
It shows a z-shaped morphology at $\lambda$6 cm, has extended Ly$\alpha$-emission, 
which is roughly aligned with the radio jet axis \citep{1990ApJ...365..487M}, 
is embedded in extended diffuse X-ray emission indicative of 
a high-density plasma region \citep{2003MNRAS.341..729F}, and is surrounded
by an apparent overdensity of faint red galaxies \citep{2003MNRAS.341L..55T}. 

Due to its proximity to a bright star useful for adaptive optics (AO) imaging,
\object{3C~294} has been intensively studied using the Keck, Subaru, and Canada France Hawaii (CFHT) telescopes.
High-resolution H and K images of the \object{3C~294} system have been presented and
discussed by \citet{1999ApJ...519L.131S}, \citet{2001ApJ...556..108Q}, 
and \citet{2002ApJ...569..611S}. There is common agreement across all studies
that the \object{3C~294} morphology can best be described by two or three
components separated by  $\leq$1\arcsec, indicative of a merging system. 
It is unclear, however, which of the 
components coincides with the location of the radio source, mostly due to 
the uncertain position of the reference star used for the astrometry. 
In addition, some of the components are compact, others extended, 
adding more uncertainty to a unique assignment of the counterpart to the 
radio source.
The situation became even more complex after an analysis of archival Chandra data
of \object{3C~294} by \citet{2004ApJ...600..626S}, who found the central X-ray emission
to be better represented by two point sources. They argued that \object{3C~294} hosts
a dual AGN, one unabsorbed system associated with a compact component and 
one absorbed system associated with one of the extended components. 

Small separation (a few kiloparsec) 
bound dual AGN are predicted to be rare in general (see \citet{2019MNRAS.483.2712R}).
On the other hand, as discussed in 
\citet{2012ApJ...746L..22K}, the percentage of dual AGN can be up to 10\%,
but they are difficult to detect in particular at high redshift due to their small projected separation. In fact, only a few high-redshift (z $>$ 0.5) dual AGN are known \citep[and references therein]{2018A&A...610L...7H}.
If 3C 294 were to evolve to a present day massive central cluster 
galaxy, one would assume that its mass would grow mostly via major mergers in
the hierarchical Universe. Since supermassive black holes seems to reside
at the centers of all massive galaxies \citep{2013ARA&A..51..511K}, one could
expect that 3C 294 hosts a dual AGN. Thus a confirmation would be an important detection.

To unambiguously determine the properties of the \object{3C~294} system and in particular
to test whether it is a  dual AGN or an AGN pair, 
we carried out high-resolution adaptive optics (AO) supported imaging in the near-infrared (NIR) 
zJHKs bands, as well as
deep optical low-resolution spectroscopy using the Large Binocular Telescope (LBT), 
the results of which are presented 
here. We note that the AO data discussed above were taken about 15 years ago.
Since then adaptive optics AO systems and NIR detectors have become much more mature and efficient.
The only spectroscopic investigation of \object{3C~294}, by \citet{1990ApJ...365..487M}, dates back more than 
25 years and was carried out using the Shane 3 m reflector. Thus a 
revisiting of the \object{3C~294} system should give a clear answer.

Throughout the paper, we assume a flat $\Lambda$CDM cosmology with ${\rm H}_0$ = 
70 km/s/Mpc and $\Omega_{\rm M}$ = 0.3. Using this cosmology, the angular 
scale is 8.45 kpc per arcsecond at z = 1.786.

\section{Observations and data reduction}

\subsection{NIR AO imaging data}

High-resolution FLAO (first light adaptive optics, \cite{2012SPIE.8447E..0UE})
supported data of \object{3C~294} were recorded in the zJHKs filters 
with the NIR imager and spectrographs LUCI1 and LUCI2 at PA = 135\degr\ during 
a commissioning run on March 20, 2016 and during a regular science run on March 20, 25, and 29, 2017.
In both LUCI instruments, we used the N30-camera, which is  optimized for AO observations.
With a scale of 0.0150$\pm$0.0002"/pixel, the field of view (FoV) offered was 30\arcsec\ $\times$ 30\arcsec. 
The observing conditions were  very good during all of the nights, with clear skies and ambient seeing of 1" or less.

The LUCI instruments are attached to the front bent Gregorian focal stations of the LBT. 
The LBT offers a wide range of instruments and observing modes. It can be operated
in monocular mode (using one instrument and mirror only), binocular mode (using identical instruments
on both mirrors), or interferometric mode (light from both telescopes combined in phase). A complete description of the current
instrument suite at the LBT and its operating mode can be found in \citet{2018SPIE10702E..05R}.
The data obtained on March 20, 2016 and March 20 and 25, 2017 were taken in monocular mode, while the data 
from March 29, 2017 were taken in binocular mode.
For the latter, the integrations and offsets were done strictly in parallel. 

The FLAO system senses the atmospheric 
turbulence by means of a pyramid-based wavefront sensor,
which samples the pupil on a grid of 30 $\times$ 30 subapertures. 
A corrected wavefront is obtained through an adaptive secondary 
mirror controlled by 672 actuators at a frequency of up to 1 kHz. 
As a reference star we used the 12th mag star U1200--07227692 from the
United States Naval Observatory  (USNO) catalog USNO-A2.0 
\citep{1998usno.book.....M}, which is just 10" southwest of \object{3C~294}.
Given its brightness, the FLAO system corrected 153 modes with a frequency of 625 Hz.
At this distance of the target from the reference star and with this FLAO configuration, 
a Strehl ratio of about 30-40\% is expected in H and Ks bands \citep{2018SPIE10702E..0BH}. 
U1200--07227692 is a binary star with 0.135" separation and a brightness ratio of 1:1.6 \citep{2001ApJ...556..108Q},
but this did not affect the performance of the FLAO system. 

Individual LUCI exposures were one minute each, consisting of six images of 10 sec that were summed before saving. 
On any given night, integrations between 7 and 37 min in total were taken in one filter before moving to the next.  
 Table \ref{nirdates} gives a log of the observations.  
Between each one-minute exposure the telescope was shifted randomly within a 2\arcsec\ by 
2\arcsec\ box to compensate for bad pixels. Since the bright reference star is the only object in 
the field present in one-minute integrations, larger offsets would not have made any difference
as the detector is particularly prone to persistence. The small offsets made sure that none
of the images of \object{3C~294} fell on a region on the detector affected by persistence from the
bright reference star. In Table 
\ref{nirsum} a breakdown of the total integration times by filter and instrument is given.

\begin{table}[h]
\centering
\vspace*{.2cm}
\begin{tabular}{l|cccc}
\hline
Date & Instrument & Filter & N$_{\rm images}$ & T$_{\rm int}$ [sec] \\
\hline
Mar 20 2016 & LUCI1 & Ks & 22 & 1320 \\
\hline
\multirow{2}{*}{Mar 20 2017} & \multirow{2}{*}{LUCI2} &H & 7 & 420 \\
& & Ks & 37 & 2220\\
\hline
\multirow{3}{*}{Mar 25 2017} & \multirow{3}{*}{LUCI1} &z & 31 & 1860 \\
& & J & 30 & 1800  \\
& & Ks & 32 & 1920 \\
\hline
\multirow{6}{*}{Mar 29 2017} & \multirow{3}{*}{LUCI1} &J & 15 & 900 \\
& & H & 30 & 1800 \\
& & Ks & 19 & 1140 \\
& \multirow{3}{*}{LUCI2} &J & 15 & 900 \\
& & H & 30 & 1800 \\
& & Ks & 20 & 1200 \\
\hline
\end{tabular}
\caption[]
{Breakdown of the observations by date, instrument, and filter.}
\label{nirdates}
\end{table}

\begin{table}[h]
\centering
\vspace*{.2cm}
\begin{tabular}{cccc}
\hline
Filter & LUCI1 [s] & LUCI2 [s] & T$_{\rm total}$ [s]\\
\hline
z & 1860 & - & 1860\\
J & 2700 & 900 & 3600\\
H & 1800 & 2220 & 4020\\
Ks & 4380 & 3420 & 7800\\
\hline
\end{tabular}
\caption[]
{Total integration times per filter and instrument and for both instruments combined.
Combined exposure times range from 31 minutes to over 2 hours.}
\label{nirsum}
\end{table}

The data were first corrected for non-linearity using the prescription given in
the LUCI user manual, then sky-subtracted and 
flat-fielded. Sky-subtraction was achieved by forming a median-combined 
two-dimensional sky image out of all data in one filter set, then subtracting a 
scaled version from each individual exposure. Given the fine sampling of 0\farcs015/pixel 
scale, we saw only about 10 counts/sec in the H and Ks bands.  With such low backgrounds 
there would have been no benefit in using a running mean or a boxcar for the sky 
subtraction. Flatfields were created out of sets of higher and lower background twilight 
images taken at zenith, which were separately median-combined after scaling them, subtracted 
from each other, and normalized. Finally, the images were corrected for bad pixels
by linear interpolation.
A bad pixel mask was created out of the highly exposed flatfields to identify
cold pixels and dark images to identify hot pixels.

\begin{figure}[ht]
 \centering
\includegraphics[width=.24\textwidth]{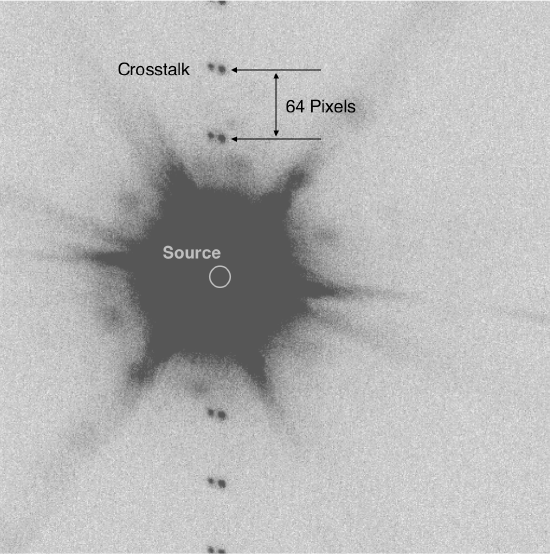}
\includegraphics[width=.24\textwidth]{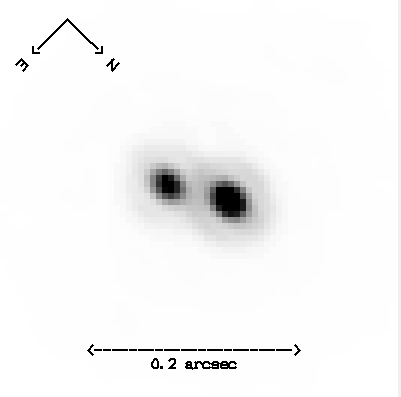}
\caption{Left) Image and channel crosstalk images of the reference star U1200-07227692. The center of the reference star is indicated as source. These crosstalk images have separations of exactly 64 pixels, the width of each of the 32 
parallel amplifier channels of the LUCI HAWAII2-RG detector.  
The well-separated images of the individual components of the reference star are remarkable.
Right) Logarithmic image of the center of the reference star to show the two components 
separated by 0\farcs135.
The Ks-band image is taken at PA = 135\degr.}
\label{xtalk}
\end{figure}

The most difficult part of the data reduction was the stacking of the images. Except for 
the saturated AO reference star and the barely visible radio galaxy, no further objects are present on the reduced images 
that could be used for the alignment of the images. We thus explored three alternative
possibilities for the alignment: a) to use the world coordinate system (WCS) information given in the image headers; 
b) to use a two-dimensional fit to the centers of both saturated  
components of the reference star after masking the saturated pixel at their
centers; and c) to take advantage of the channel crosstalk shown by the detector, which
leaves a non-saturated imprint of the reference star in every channel of the detector 
on the frame (see Fig. \ref{xtalk}).

The simple approach using the WCS information failed, likely because of residual flexure within LUCI, 
resulting in a visibly "smeared" image of the reference star. We thus did not pursue this approach further.
Each of the two alternative methods has its advantages and disadvantages. Determining a centroid of 
a saturated core leaves some uncertainty but it benefits from a high signal-to-noise ratio (S/N) in its outer parts.
The individual channel crosstalk images have a lower S/N, but combining 10-15 of them from adjacent channels
increases the signal considerably. 
We tested both methods using a data set taken in the Ks filter.
The resulting offsets agree within 1/10 of a pixel. Given that we opted for integer pixel shift before 
combining the images, both methods delivered equally good results for our purposes. In the end
we decided to use the offsets derived from the centroids to the cores of the two components of the reference star.

The aligned images were first combined per filter, instrument, and night, 
then per filter and telescope, and finally per filter. The relative orientation of the
detector on the sky between the two instruments differs by less than one degree. 
In addition, the pixel scale between the N30 cameras in LUCI1 and LUCI2 differs by less 
then $10^{-4}$. Thus no rotation or rebinning was applied before combining the 
data sets from the different instruments. 

\subsection{Optical spectroscopy}

\begin{figure}[t]
 \centering
\includegraphics[width=.49\textwidth]{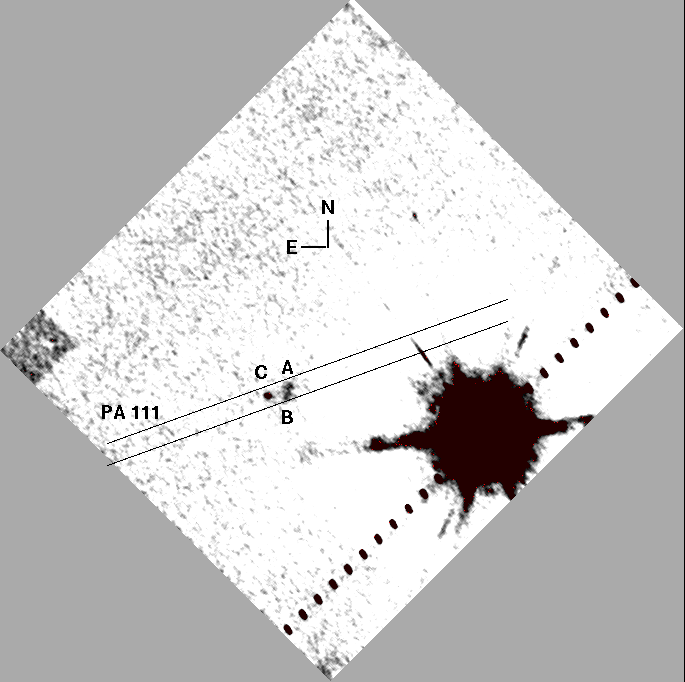}
\caption{Orientation of the slit with respect to the \object{3C~294} system for the 
MODS spectroscopy.}
\label{figslit}
\end{figure}

Optical longslit spectra of the \object{3C~294} system were taken on the night of May 11-12, 2016 using the 
multi-object double CCD spectrographs MODS1 and MODS2 \citep{2010SPIE.7735E..0AP} in homogeneous binocular mode. 
The MODS instruments are attached to the direct Gregorian foci at the LBT.
The target was observed at PA = 111\degr\ in 
order a) to have all components of the \object{3C~294} system in the slit, b) to see whether components a, b, and c are at the
same redshift, and c) to minimize the impact of the nearby bright star on 
the data quality (see Fig. \ref{figslit} for the configuration of MODS1/2 and Fig. \ref{3c294imafig}
for more details on the components). To do so, we performed a blind offset from the 
bright reference star, with the offset determined from the AO K-band image. The
expected accuracy of the positioning is $\sim$ 0\farcs1.
We used a 1" slit and the gratings G400L for 
the blue and G670L for
the red channel, giving a spectral resolution of about 1000
across the entire optical band. Integration times
were 3 $\times$ 1200 sec. Observing conditions were not photometric with variable cirrus,
but excellent seeing ($\sim$ 0\farcs7 full width at half maximum (FWHM)).

The basic data reduction (bias subtraction and flatfielding) was carried out using
the modsCCDRED-package developed by the MODS team \citep{richard_pogge_2019_2550741}. 
The extraction of one-dimensional spectra was carried out using the standard 
image reduction and analysis facility (IRAF) 
{\em apall} task. As the spectra of \object{3C~294} did not show any continuum (Fig. \ref{3c294spectra}) 
and were taken through variable cirrus, we did not carry
out a spectrophotometric flux calibration. Wavelength calibration was done using spectral images
from calibration lamps and verified using the night sky emission lines. The 
resulting accuracy was $\sim$ 0.1\AA\ rms. The resulting spectra were first averaged 
per telescope and channel and then combined per channel.

\section{Results}

\subsection{AO NIR images of \object{3C~294}}

\begin{figure*}[htbp]
 \centering
\includegraphics[width=0.25\textwidth]{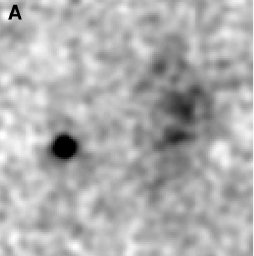}
\includegraphics[width=0.25\textwidth]{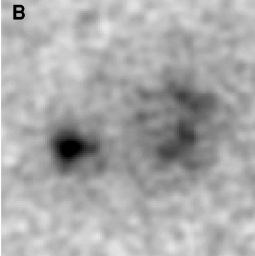}
\includegraphics[width=0.25\textwidth]{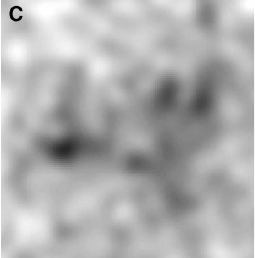}\\
\vspace*{.2cm}
\includegraphics[width=0.25\textwidth]{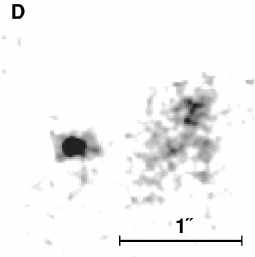}
\includegraphics[width=0.50\textwidth]{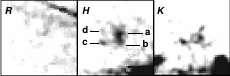}\\
\vspace*{.5cm}
\includegraphics[width=0.29\textwidth]{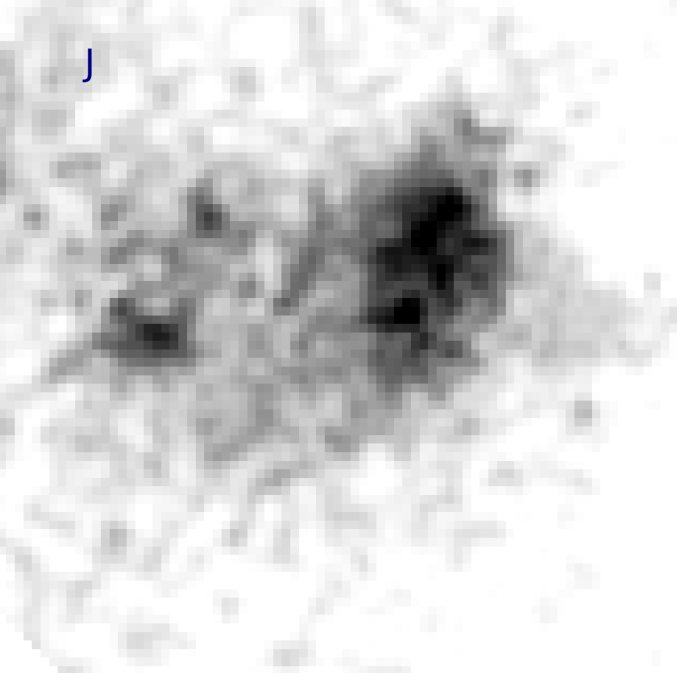}
\hspace*{.1cm}
\includegraphics[width=0.29\textwidth]{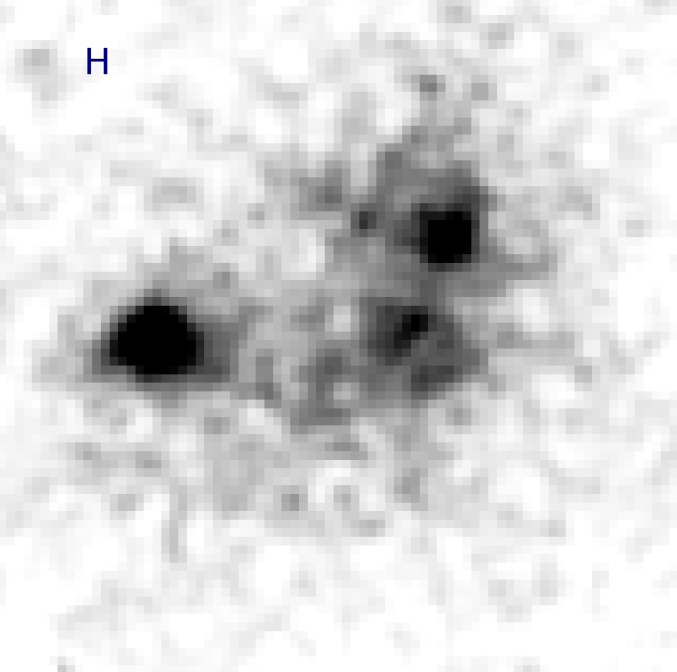}\\
\vspace*{.1cm}
\includegraphics[width=0.29\textwidth]{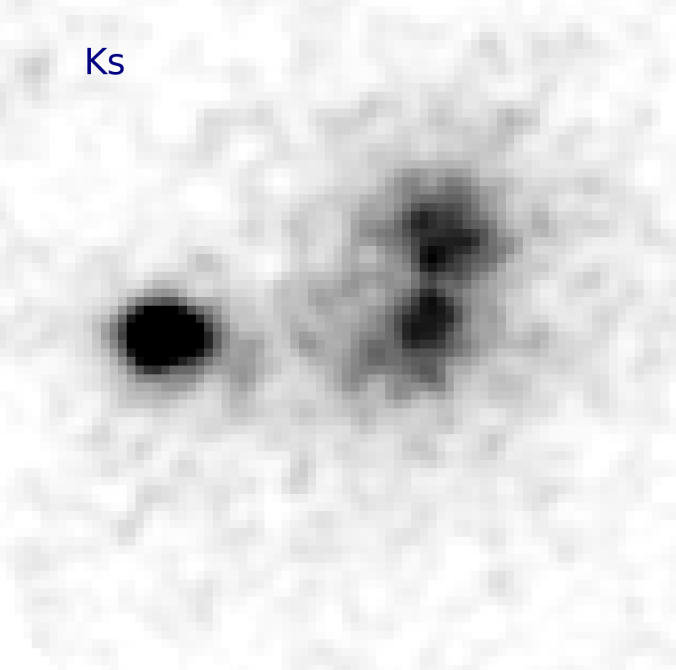}
\hspace*{.1cm}
\includegraphics[width=0.29\textwidth]{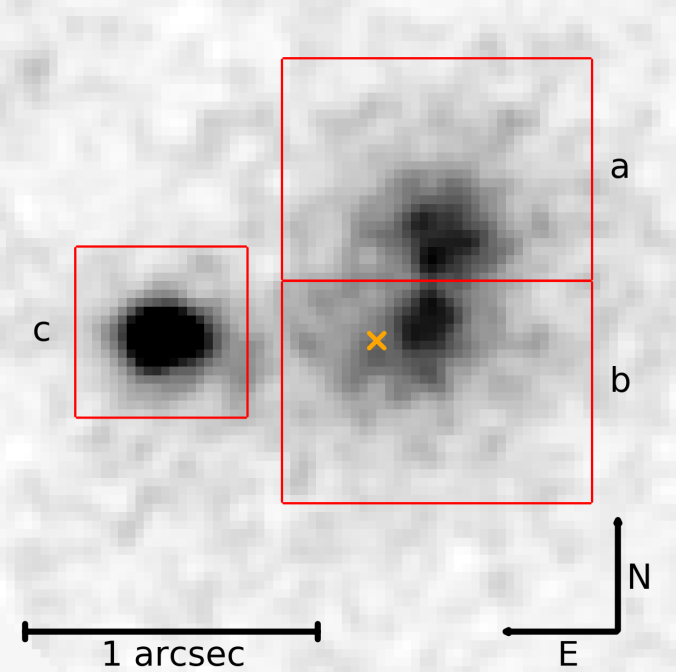}\\
\caption{Two uppermost rows)  AO NIR images of 3C 294 from earlier
publications. A) Keck\,II NIRSPEC K' image from \citet{2004ApJ...600..626S}, B) Subaru IRCS K' image also from 
\citet{2004ApJ...600..626S}, C) CFHT Hokupa K' image from
\citet{1999ApJ...519L.131S}, and D) Keck II SCam H-band image from
\citet{2001ApJ...556..108Q}.  The images A-D are adapted from \citet{2004ApJ...600..626S}. The three images right of image D show
R (HST), H, and K CFHT PUEO data from \citet{2002ApJ...569..611S}.
Two lowest rows)
Our LBT FLAO and LUCI JHKs-images of \object{3C~294}. The fourth image shows the scale, orientation, and apertures for the components used for the photometric analysis. 
They are 36 $\times$ 26 pixel (0\farcs54 $\times$
0\farcs39) for components 
a and b and 20 $\times$ 20 pixel (0\farcs3 $\times$ 0\farcs3) for component c. The 
labeling follows \citet{2002ApJ...569..611S}.
The cross marks the position of the radio core, its size reflects its positional error.
}
\label{3c294imafig}
\end{figure*}

The final AO J-, H-, and Ks-band images are shown in the lower part of 
Fig. \ref{3c294imafig}. 
They have been binned by a factor of two to emphasize structures 
more clearly. 

As discussed in \citet{2001ApJ...556..108Q} and
\citet{2002ApJ...569..611S}, 
\object{3C~294} can be separated into two main components separated by about 1\arcsec: a 
compact core-like structure to the east and a structure elongated north--south to the west.
The elongated structure seems to consist of two knotty components also separated by roughly 1\arcsec. 
No emission from \object{3C~294} was detected in the z-band image. This  is
probably due to the shallow depth of the image, the much lower Strehl ratio,
and/or the increasing extinction compared to the redder bands. We note that \object{3C~294} has only barely been 
detected in optical broadband images with the Hubble Space Telescope (HST)  (m$_{\rm R}$ = 23.4$\pm$0.8, \cite{2002ApJ...569..611S}).

Contrary to earlier observations, the two components of the western component
are clearly separated.  A comparison of earlier H- and K-band images with our data
is shown in Fig. \ref{3c294imafig}. 

We do not see a clear separation of the western structures in the J band.
The reason for that is not clear. It cannot be due to extinction by dust 
as the H and Ks bands would be less affected by that 
(Ks band probes the rest-frame wavelength at $\sim$ 7700 \AA\ and J band probes the rest-frame wavelength at $\sim$ 4400 \AA). It is more likely due to the 
lower Strehl ratio, which is expected to drop by 10-20\% 
between the H band and the J band. We do not detect the component d north of component c
discussed in \citet{2002ApJ...569..611S} in any of our images.  We should have detected it in the 
 H band as both components have a similar brightness  in this filter
 according to \citet{2002ApJ...569..611S}.  
 Thus, feature d is either a transient phenomenon or is not physical, that is, it is a statistical noise fluctuation.
 There may, however, be some (very blue?) emission northwest of component c in the J-band image,
 which is not at the same location as component d.

The core-like component c appears slightly extended in an east--west direction on the images. 
This can also be seen in data from other telescopes shown in Fig. \ref{3c294imafig}.
Its measured FWHM is 0\farcs31 $\times$ 0\farcs20 in the H band and 0\farcs23 $\times$ 0\farcs17 in the
Ks band. If component c were a pure point source, we would expect a FWHM of about 
0\farcs08 at 10\arcsec\ from a 12th mag reference star \citep{2018SPIE10702E..0BH}.
Interestingly, the major axis of the elongation is perpendicular to the one seen for the reference star 
in Fig. \ref{xtalk}. The latter is most likely due to 
vibrations of the telescope (D. Miller, priv. com.) and present on some of the individual 
images. The former could be due to tilt anisoplanatism as it is along the axis 
joining the reference star and component c. \citet{1994JOSAA..11..368O} derived a 
formalism to estimate the tilt anisoplanatism for the Keck telescope. Using their estimates
and scaling them to the LBT (8.4 m) and wavelength (Ks, 2.15 $\mu$m), we would expect a tilt anisoplanatism
of 0\farcs022 and 0\farcs019 in x and y-direction, respectively. We would thus expect the FWHM of component c to
be on the order of 0\farcs10 - 0\farcs12 with an axis ratio of $\sim$ 1.2. This is much smaller then what
we measure. We thus believe that component c is most likely not stellar. 

This is in contrast to \citet{2001ApJ...556..108Q} who
found component c to be unresolved. Unfortunately, due to the small separation of the two components
of the reference star (0\farcs135) and their saturation in the core, they can hardly be used for 
comparison. A formal Gaussian fit to each of the two components of the reference star ignoring the 
central saturated pixels (22 pixels for the brighter and 11 pixels for the fainter component)
results in a FWHM of $\sim$ 0\farcs08.

\begin{table}[h]
\centering
\begin{tabular}{c|llll}
\hline  
Comp     & J & J - H & J - Ks & H - Ks\\
\hline
a     & 21.81$\pm$0.16 &  1.00$\pm$0.19 &  1.93$\pm$0.18 &  0.93$\pm$0.14 \\
b     & 21.83$\pm$0.16 &  0.93$\pm$0.19 &  1.83$\pm$0.18 &  0.90$\pm$0.14 \\
c     & 22.53$\pm$0.13 &  1.16$\pm$0.16 &  2.07$\pm$0.15 &  0.93$\pm$0.11 \\
Total & 20.48$\pm$0.31 &  1.05$\pm$0.35 &  2.02$\pm$0.34 &  0.97$\pm$0.22 \\
\hline
\end{tabular}
\caption{JHKs photometry of \object{3C~294} and its components. 
\label{3c294phot}}
\end{table}

\begin{table*}[h]
\centering
\begin{tabular}{l|ccccc}
\hline  
Filter     & J & H & K' & Ks & K\\
\hline
LBT                         & 20.48$\pm$0.31 & 19.43$\pm$0.17 & - & 18.56$\pm$0.14& -\\
\citet{1990ApJ...365..487M} & -              & -              & - & - & 18.0$\pm$0.3\\
\citet{1999ApJ...519L.131S} & -              & -              & 18.3$\pm$0.3 & - & - \\
\citet{2001ApJ...556..108Q} & -              & 19.4$\pm$0.2   & 18.2$\pm$0.2 & - & - \\
\citet{2002ApJ...569..611S} & -              & 18.2$\pm$0.3   & - & - & 17.76$\pm$1.0 \\
\citet{2003MNRAS.341L..55T} & 19.53$\pm$0.30 & 18.64$\pm$0.27 & - & 17.78$\pm$0.07& -\\ 
\hline
\end{tabular}
\caption{Comparison of JHK photometry of the entire \object{3C~294} system. To derive 
the magnitudes from \citet{2002ApJ...569..611S}, the fluxes from components
a, b, and c have been summed.
\label{litphot}}
\end{table*}

Judging from the images, component c seems to be redder than components a and b. To verify 
this we performed aperture photometry on the individual components using the apertures indicated in 
Fig. \ref{3c294imafig}. Calibration was done using star P272--D from \citet{1998AJ....116.2475P}; the data have
been corrected for galactic extinction following \citet{2011ApJ...737..103S}.
It is $\leq$0.01 mag in all bands. 

The results shown in Table \ref{3c294phot} indicate that 
components a and b have similar brightnesses and colors, while component c is about 0.7 mag
fainter and about 0.2 mag redder. Photometry of the entire system has been presented
by \citet{1990ApJ...365..487M}, \citet{1999ApJ...519L.131S}, \citet{2001ApJ...556..108Q},
\citet{2002ApJ...569..611S}, and \citet{2003MNRAS.341L..55T}.
A comparison to our data is shown in Table \ref{litphot}. 
There is a wide spread in the photometry, with differences of up to $\sim$ 1 mag in particular to 
\citet{2003MNRAS.341L..55T}. It is not clear where the differences in the photometry 
come from. Unfortunately, the size of the aperture used is not always given.

Photometry of individual components have been derived by \citet{2001ApJ...556..108Q} and 
\citet{2002ApJ...569..611S}. A comparison to our data is shown in Table \ref{litcompphot}. 
The difference between the \citet{2001ApJ...556..108Q} and \citet{2002ApJ...569..611S} data 
is large, approximately two magnitudes, while our measurements are somewhere in between.
Again, the size of the aperture used is not given for either the Keck-data or CFHT-data.

We also created J-H, H-Ks, and J-Ks color maps for \object{3C~294} to search for any evidence of 
spatially-dependent extinction in the three components. No obvious feature was detected. 

\begin{table*}[h]
\centering
\begin{tabular}{l|ccc|ccc}
\hline  
Filter     & \multicolumn{3}{c}{H} |& \multicolumn{3}{c}{K} \\
\hline
Component & a & b & c & a & b & c \\
\hline
LBT                         & 20.88$\pm$0.16 & 20.83$\pm$0.11 & 21.37$\pm$0.09 & 20.04$\pm$0.09 & 19.96$\pm$0.09 & 20.50$\pm$0.07 \\
\citet{2001ApJ...556..108Q} & - & - & 22.0$\pm$0.2 & - & - & 21.7$\pm$0.2\\
\citet{2002ApJ...569..611S} & 19.0$\pm$0.1 & 19.5$\pm$0.2 & 20.2$\pm$0.2 & 18.7$\pm$0.4 & 18.4$\pm$0.4 & 19.3$\pm$0.8\\
\hline
\end{tabular}
\caption{Comparison of HK photometry of components a, b, and c of the \object{3C~294} system. The K images have been taken through 
different filters (see Table \ref{litphot}) but the differences are small and did not affect the large differences
seen between the individual measurements.
\label{litcompphot}}
\end{table*}

\subsection{Optical spectra of the \object{3C~294} system}\label{modsspec}

\begin{figure*}[ht]
 \centering
\includegraphics[width=.485\textwidth]{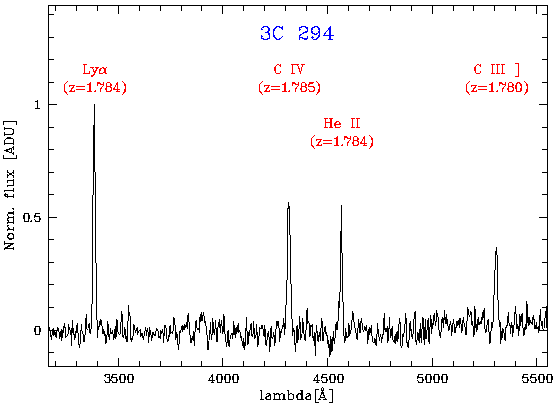}
\includegraphics[width=.485\textwidth]{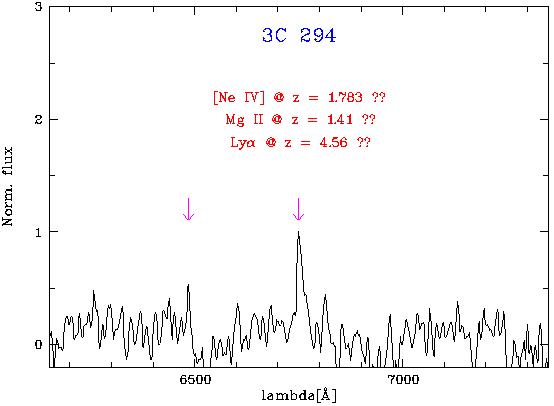}\\
\includegraphics[width=1\textwidth]{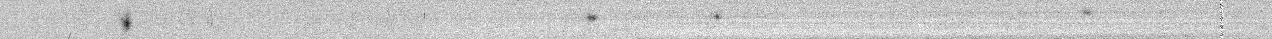}\\
\includegraphics[width=1\textwidth]{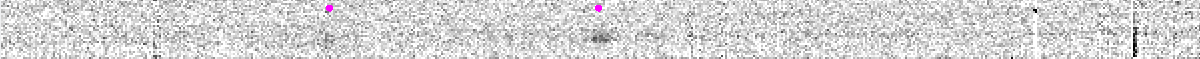}\\
\caption{Top) Our one-dimensional MODS spectra of the \object{3C~294} system in the 
blue (left) and (red)
channel with the line identifications. The positions of the two emission features 
at $\lambda$6485 and $\lambda$6749.4 \AA\ discussed in the text are indicated by arrows.
Bottom) Our two-dimensional MODS spectrum of the 3C 294 system in the blue channel (upper panel) showing the 
four emission lines as well as an excerpt of the two-dimensional spectrum in the red channel (lower panel)
showing the emission line detected at $\lambda$6749.4 \AA. The blue spectrum is shown across its full spectral range, the red spectrum with an identical width then the one-dimensional spectrum centered 
on the line at $\lambda$6749.4 \AA. The two emission features are labeled by magenta dots.}
\label{3c294spectra}
\end{figure*}

The one-dimensional and two-dimensional spectra of the \object{3C~294} system are 
shown in Fig. \ref{3c294spectra}. Despite our 2 hr integration time, we do
not detect any obvious continuum. When rebinning the spectra in spectral and spatial 
direction by a factor of five, a hint of continuum can be glimpsed in the red channel. 
However, with this rebinning, the potential continuum of components a, b, and c would 
overlap, preventing us from distinguishing between the two components (see below).
We detected the Ly$\alpha$ $\lambda$1215 \AA, C IV $\lambda$1550 \AA,
He II $\lambda$1640 \AA,\ and C III$]$ $\lambda$1907/1909 \AA\ lines in the blue channel.
The C III$]$ line is blue-shifted with respect 
to the other lines. This is not unusual in AGN
(e.g., \citet{2012ApJ...744....7B})
and may be a result of the increased intensity of the transition at $\lambda$1907 \AA\ relative to 
the one at $\lambda$1909\ \AA\ at low densities \citep{1981ApJ...249...17F}.

Only one emission line at $\lambda$6749.4\ \AA\ can be detected in the red channel. It
is faintly present in the individual spectra from both MODS instruments and is not an 
artifact of the data reduction. Under the assumption that it originates from the 
same source as the emission lines in the blue channel, we identify this line
with the Ne $[$IV$]$ doublet $\lambda$2424/2426 \AA\ at z = 1.783. 
Similarly to our C III$]$ $\lambda$1907/1909 \AA\ line in the blue channel
we would not be able to separate the double lines individually given an instrumental resolution
of $\sim$ 5\AA. If the identification is correct,
we may even associate the very faint emission at $\lambda\sim$ 6485 \AA\  with C II$]$ 
$\lambda$2326 \AA\ at z $\sim$ 1.788. The position of this line coincides with a forest
of night-sky emission lines, however, so this detection is tentative.
To our surprise, we do not detect Mg~II
$\lambda$2798 \AA\
in our spectra, which we would expect at $\lambda\sim$ 7790 \AA\ . However, Mg~II is
relatively weak in radio galaxies \citep{1993ARA&A..31..639M} and line ratios
can vary strongly in high-redshift radio galaxies \citep{2008MNRAS.383...11H}.

Ignoring the redshift determination for the C III$]$ $\lambda$1907/1909 \AA\ line
and the uncertain identification of the C II$]$ 
$\lambda$2326 \AA\ line, our redshift for \object{3C~294} is z = 1.784$\pm$0.001. 
This is somewhat lower then the redshift z = 1.786 quoted
by \citet{1990ApJ...365..487M}, who did not quote errors. With their instrumental resolution 
of $\sim$ 12 - 15\AA, an error for their redshift $\delta$ z of at least  0.001 is a fair assumption. Thus
our redshifts for 3C 294 agree within the errors. We will use z = 1.784 as the redshift for 3C 294 for the remainder of the paper.

Similarly to \citet{1990ApJ...365..487M}, we see a spatially extended (6") Ly$\alpha$ line.
Unfortunately, since the spectra of components a and b overlap in dispersion 
direction, we cannot probe any dynamics using this line.

The absence of a continuum poses the problem of a unique assignment of the 
emission line at $\lambda$6749.4 \AA\ to one of the components of \object{3C~294}. To derive this, 
we did the following: we first fitted a slope through the centers of the emission
lines in the blue channel except for the spatially extended Ly$\alpha$ line.
We then did the same through the trace in both channels of a well-exposed star 
observed on the same night to infer the spatial offset. Afterwards we determined the 
spatial offset between the trace of the star in the blue and the red channels to 
finally estimate the expected position of the trace of either component in the red 
channel at $\lambda$6749.4 \AA. It turned out that the line emission at $\lambda$6749.4 \AA\
is about 4$\pm$2 pixels above the 
expected trace of components a and b and about 3$\pm2$ pixels below the expected position
of component c. Thus we cannot unambiguously tell whether the line at $\lambda$6749.4 \AA\ originates
from the same source as the other UV lines and thus assign it to
components a and/or b or c.

\section{Discussion}

\subsection{Which component is associated with the radio source?}

There has been no common agreement as to which of the components of the \object{3C~294} system
is indeed the NIR counterpart of the radio source. Using a position of the 
radio source of $\alpha$,$\delta$ = $14^{\rm h} 06^{\rm m} 44\fs074\pm0\fs005, +34^{\degr} 25' 40{\farcs}00\pm0\farcs05\
(2000)$\ based on \citet{1990ApJ...365..487M}, \citet{2004ApJ...600..626S} found
it practically coincident with component b. On the contrary, \citet{2001ApJ...556..108Q} 
associated the radio source with component c. Their positions of the radio core and the
reference stars differed by +0\farcs07 and +0\farcs5 from the ones used by \citet{2004ApJ...600..626S},
respectively. The main difference for the position of the
reference star comes from the sources used (HST FGS observation
by \citet{2004ApJ...600..626S} and the USNO--A2.0 catalog by
\citet{2001ApJ...556..108Q}). 
It is surprising that \citet{2001ApJ...556..108Q} associated the radio
source with component c although they found a smaller separation between the 
reference star and the radio source than \citet{2004ApJ...600..626S}. 
As can be seen from Fig. \ref{figslit}, component b is closer to the reference star than component c.

The most accurate position for the reference star comes from the Gaia Data Release 2 
(DR2) \citep{2016A&A...595A...1G,2018A&A...616A...1G}, which gives $\alpha$,$\delta$ = 
$14^{\rm h} 06^{\rm m} 43\fs356, +34^{\degr} 11' 23{\farcs}273$ (2000) with an 
error of about 0.05 mas for $\alpha$ and  $\delta$ each. The proper motion is 
$\alpha$,$\delta$ = +8.80,1.66 mas/yr.
Since the spatial resolution of the Gaia DR2 is about 0\farcs4, the binary star is not 
resolved and we can assume that the position given above refers to the center of light.
This position is in very good agreement with the one determined by 
\citet{2004ApJ...600..626S}.

Using the coordinates of the reference star from the Gaia DR2 and the 
radio coordinates of \object{3C~294} from \citet{2004ApJ...600..626S}, we can now
predict the position of the radio core of \object{3C~294} on our AO NIR images.
We used the crosstalk images of the reference star to predict the position
of the center of light of the binary star with an accuracy of $\pm$1 pixel 
($\pm$0\farcs015).
The result is shown in Fig. \ref{3c294imafig}. As in \citet{2004ApJ...600..626S},
our result indicates that the NIR counterpart to the \object{3C~294} radio source is
component b. The overall error budget (accuracy of center of light for the reference
star in our images, errors of the positions for the reference star and the radio source)
does not exceed 0\farcs1, meaning that component c can be ruled out as the NIR counterpart
with high confidence.

\subsection{Nature of the NIR counterpart of \object{3C~294}}

We are now convinced that (at least) component b is the NIR counterpart
to the radio source. We also know that extended redshifted Ly$\alpha$ emission 
centered on the radio core has been detected \citep{1990ApJ...365..487M}. In addition, since 
our spectroscopic results agree well with \citet{1990ApJ...365..487M}, 
the redshift of z = 1.784 for \object{3C~294} is now solid. There is agreement that components
a and b show a non-stellar morphology.  These components are clearly separated 
in our H- and Ks-band data. Given our spatial resolution and the distance of \object{3C~294}, 
we cannot decide whether components a and b represent 
two galaxies in the process of merging, whether they correspond to a single 
galaxy with an absorption feature along the line of sight, or whether they are two 
galaxies at different redshifts. 

Surprisingly, both components have similar brightnesses and colors. At z = 1.784,
our JHKs images correspond roughly to rest-frame BVI-data. With B-V $\sim$ 1.0,
components a and b seem to be dominated by late stars of type K and M. If we assume
that components a and b are galaxies, we obtain M$_{\rm K}$ $\sim$ -25.3
from their Ks-band magnitudes of $\sim$ 20.0. This includes a lower limit for the K-band correction 
of K = -0.5 (which is between
K = 0 and -0.5 depending on galaxy type; \citet{1997A&AS..122..399P,2001MNRAS.326..745M})
at z = 1.784. No evolutionary correction has been applied. We note that this is an
upper limit for the host galaxy of \object{3C~294} as we do not know the contribution of 
the active nucleus to the total flux. If 90/50/10\% of the flux from component b 
is from the host galaxy, we would derive M$_{\rm K}$ $\sim$ -25.2/-24.6/-22.8.
This is between 2 mag brighter and 0.4 mag fainter 
than a M$_{\rm K}^{\ast}$ galaxy \citep{2017MNRAS.465..672M}.

\subsection{Nature of component c}\label{disccompc}

What is the nature of component c? \citet{2004ApJ...600..626S} discussed the intriguing 
possibility that \object{3C~294} hosts two active nuclei. This idea stems from an analysis of
archival Chandra data, where the X-ray emission from \object{3C~294} could better be described by
a superposition of two point sources. Based on the X-ray/optical flux ratio, they argued that
component c is unlikely to be a foreground galactic star but could well host a second
active nucleus. We found component c about 0.8 mag brighter 
than \citet{2001ApJ...556..108Q}, but even then its X-ray/optical flux ratio of $\sim$ 1.1
indicates that the source is rather an AGN then a star using the arguments of 
\citet{2004ApJ...600..626S}. If it is a star, it could be a carbon star. These stars
are often found in symbiotic X-ray binaries (e.g., \cite{2014ApJ...780...11H}) and
have very red V-K and H-K colors similarly to what we found \citep{2001ApJ...558..309D}.
However, with V-K $\sim$ 5, and M$_{\rm V}$ $\sim$ -2.5 typical for carbon stars 
\citep{1998A&A...338..209A}, the distance modulus would place our "star" well outside 
our Galaxy. In addition, component c appears extended in our data supporting an 
extragalactic origin.

Unfortunately, the results from our spectroscopy are of little help.
We now know that the radio core coincides with component b. It is thus
reasonable to assume that the UV lines detected in the blue channel, which have 
also be seen by \citet{1990ApJ...365..487M}, originate from that region.
We note that \citep{1990ApJ...365..487M} used a 2" wide slit at PA = 200\degr,
which means that their spectra of components a, b, and c overlapped in 
spectral direction. As discussed in Sect. \ref{modsspec}, we cannot unambiguously
assign the emission line at $\lambda$6749.4 \AA\ to component b or c based
on its spatial location. Given its spectral position one can reasonably assume
that this line originates from the same region as all the other UV lines, namely
from component b. If this is the case, the nature of component c remains a mystery.

Although speculative, we briefly discuss the consequences if the emission line at $\lambda$6749.4 \AA\ 
belongs to component c. One exciting alternative 
would be that the line originates from the Ne $[$IV$]$ doublet 
$\lambda$2424/2426 \AA\ at z = 1.783 from this component. 
This would make \object{3C~294} indeed a dual, perhaps bound AGN separated by a few kiloparsec
as discussed by 
\citet{2004ApJ...600..626S}. However, they speculated that the AGN coincident 
with component c  is less powerful but does not suffer so much from extinction. 
In that case one would expect to see the UV lines (in particular Ly$\alpha,$ which is typically
a factor of $\sim$60 stronger than Ne $[$IV$]$ $\lambda$2424/2426 \AA\
in radio galaxies \citep{1993ARA&A..31..639M}) in the blue channel, unless 
they are unusually weak. An inspection of the two-dimensional
spectrum in the blue channel did not reveal any second component in spatial direction.
Thus we do not have strong support for the dual AGN scenario based on our spectroscopy.

If not at the same redshift, component c could be at a different redshift.
Given the faintness of the optical counterpart and emission line, and showing X-ray emission, it most likely 
originates from an AGN. 
Judging from the composite AGN spectrum of \citet{2001AJ....122..549V},
the most prominent lines in the optical are H$\alpha$, H$\beta$, the [O II, O III] lines, 
and Mg~II. Out of these, it cannot be H$\alpha$ at z = 0.029, because its
NIR luminosity would be much too low unless it suffers from extreme absorption. 
It also cannot be 
H$\beta$ at z = 0.38 because then we should have seen H$\alpha$ at $\lambda$9054 \AA,\, which is
normally much stronger, and/or the [O II] or [O III] lines at $\lambda$3727 and $\lambda$5007 \AA, respectively. 
The same argument applies for the [O III] line at z = 0.35. 
The [O II] line at z = 0.81 would be an interesting possibility.  The typically much 
stronger H$\beta$ and O [III] lines would be shifted towards $\lambda$9000 \AA, where
there is a strong forest of night-sky  emission lines. However, one would 
then easily see Mg~II $\lambda$2798 \AA, which is normally also stronger then [O II].
Thus, Mg~II $\lambda$2798 \AA,\, which would be at z = 1.41, remains as the most reasonable 
line identification. All optical emission lines redward of Mg~II are redshifted beyond $\lambda$9000 \AA\ 
and are thus hard to detect or are out of the optical range. Only the C IV and C III] UV lines remain. 
These would be redshifted to $\lambda$3735 and $\lambda$4600 \AA, respectively, but are
not present in our spectra. These lines are often faint or not present in type II 
quasi stellar object (QSO) candidates  
at high redshift \citep{2013MNRAS.435.3306A}, so it would not be surprising. 
One caveat of all of the options discussed above is that even at z = 1.41 the host 
galaxy of an AGN must substantially absorb the emission from \object{3C~294}. This has not been seen. Thus, 
even the most reasonable option (Mg~II at z = 1.41) is not convincing.

Alternatively, the emission line in component c could derive from a redshifted UV line.
The strongest UV lines in a QSO spectrum are Ly$\alpha$ $\lambda$1215 \AA, C~IV $\lambda$1549 \AA,\ and 
C~III] $\lambda$1909 \AA\ 
\citep{2001AJ....122..549V}. This would move the AGN to redshifts 
beyond z = 2.5, with \object{3C~294} then being along the line of sight to component c.  Its redshift would then 
be z = 2.54 (C III]), 3.36 (C IV), or 4.56 (Ly$\alpha$), respectively.  If this is the case, the UV lines 
should be absorbed to some extent by \object{3C~294}.

There are eight QSO at z $>$ 2.5, up to z = 5.2, in the Chandra Deep Field North 
(CDFN, \citet{2001AJ....122.2810B,2002AJ....124.1839B}). These eight sources all
share the same properties with component c. Their soft X-ray flux in the 0.5-2.0 
keV band is between 0.3 and 6 $\times 1e^{-15}$ ergs cm$^{-2}$ s$^{-1}$, their K magnitudes are $\sim$ 21, 
their V magnitudes are $\sim$ 24, and their spectroscopic signatures
include strong, broad Ly$\alpha$, sometimes also accompanied by
strong C+IV and C~III]. Given the faintness of our emission line and the
absence of a second line, it is reasonable to assume that this would
correspond to Ly$\alpha$ at z = 4.56. 

\subsection{Consequences for an AGN pair}

Our results do not allow us to discriminate between the 
dual AGN or projected AGN scenario. 
Even the latter would not contradict the interpretation by
\citet{2004ApJ...600..626S} that \object{3C~294} hosts an obscured AGN centered
at component b, while a second much fainter but not obscured AGN is
coincident with component c. One natural explanation for
the differences in the photometry from various studies summarized in 
Table \ref{litphot} is the intrinsic variability of the two AGN.

Projected AGN pairs can be used for a number of astrophysical applications. Examples
are QSO-QSO clustering and the tomography of the intergalactic medium  or the
circumgalactic medium of the QSO along the light of sight to the background QSO
\citep{2006ApJ...651...61H}. The latest compilation of projected QSO pairs can
be found in \citet{2018ApJS..236...44F}. However, the number of small-separation pairs
(a few arcsec) is very small \citep{2012AJ....143..119I,2016MNRAS.456.1595M}, and they have 
mostly be derived from searches for gravitationally lensed QSOs and all have a
wider separation ($\geq2\farcs$5) than our target.  To the best of our
knowledge, no projected AGN pair with such a small separation and large
$\Delta$ z is known at present.

Due to the close separation of our system, gravitational lensing effects 
could modify the 
apparent properties of the \object{3C~294} system. A multiply-lensed QSO image of component c
would be expected for an Einstein radius of $\geq$1\arcsec. 
Since we do not see any, this would set the upper limit of the 
\object{3C~294} host galaxy to 3 $\times$ $10^{12}$ \(M_\odot\) 
at a redshift of z = 1.784 and 4.56 for the lense and source, respectively.
As the host galaxy is certainly not
point-like, any amplification must be very low. In addition,
component c could even be subject to gravitational microlensing by stars 
in the host galaxy of \object{3C~294}. This might at least in part explain the difference in 
brightness of the \object{3C~294} system shown in Table \ref{litphot}.

\subsection{Outlook}

The analysis of our deep AO images and optical spectra of \object{3C~294} did not 
allow us to unambiguously characterize the \object{3C~294} system as
the main conclusion rests on the spatial association of the 
emission line at $\lambda$6749.4\ \AA\ with either component b or c. 
If it originates from component b, the nature of component c remains a mystery.
If it originates from component c, we have support for either the dual or projected 
AGN scenario. Whether the lines originate from component b or c can be tested by 
repeating the optical spectroscopy "astrometrically" by taking  a 
spectrum of \object{3C~294} and a bright object on the FoV simultaneously, with the latter 
showing a trace on the two-dimensional spectrum.  If the line at $\lambda$6749.4\ \AA\
belongs indeed to component c, AO-aided NIR spectroscopy is the only way 
to characterize the system due to the faintness of the system and the 
probably high redshifts involved.
Not much can be learned for \object{3C~294} itself from the ground, as at 
z = 1.784 all diagnostic optical emission lines except H$_{\gamma}$ will be 
redshifted into a wavelength range where the NIR sky is opaque. At least one of the
redshifted [O~II, O~III] or H$\alpha$,$\beta$ lines is redshifted into one of 
the JHK-windows if component c is at z = 2.54, 3.36, or 4.56. 
An unambiguous determination of the nature of 3C 294 will be possible with the NIR spectrograph
NIRspec onboard the James Webb Spacec Telescope.
With its 3" $\times$ 3" Integral Field Unit covering the wavelength range 
0.67 - 5$\mu$m, a number of diagnostic lines can be observed in a very low infrared background
devoid of opaque wavelength regions.

\begin{acknowledgements}
We would like to thank the anonymous referee for the constructive comments 
that addressed a number of important points in the paper.
We would also like to thank Mark Norris and Jesper Storm
for taking the MODS-data at the LBT for us. 
This work has made use of data from the European Space Agency (ESA) mission
{\it Gaia} (\url{https://www.cosmos.esa.int/gaia}), processed by the {\it Gaia}
Data Processing and Analysis Consortium (DPAC,
\url{https://www.cosmos.esa.int/web/gaia/dpac/consortium}). Funding for the DPAC
has been provided by national institutions, in particular the institutions
participating in the {\it Gaia} Multilateral Agreement.
This work was supported in part by the German federal department for education and research (BMBF) 
under the project numbers 05 AL2VO1/8, 05 AL2EIB/4, 05 AL2EEA/1, 05 AL2PCA/5, 05 AL5VH1/5, 05 AL5PC1/1, and 05 A08VH1. 
\end{acknowledgements}

\bibliographystyle{aa}

\bibliography{3c294}

\end{document}